\documentclass[aps,pra,twocolumn,amsmath,amssymb,bibnotes]{revtex4}

\usepackage{color,epsfig,rotating,graphicx,subfigure}

\usepackage[latin1]{inputenc}
\usepackage[english]{babel}

\usepackage{dsfont}
\usepackage{textcomp}

\renewcommand{\Re}{{\rm Re}}
\renewcommand{\Im}{{\rm Im}}

\newcommand{\ri}{{\rm i}}
\newcommand{\re}{{\rm e}}
\newcommand{\rd}{{\rm d}}

\newcommand{\rs}{{\rm s}}
\newcommand{\rp}{{\rm p}}

\begin{document}
\title{Revisiting the dipole model for TINS}

\date{\today}

\author{Florian Herz$^1$, Zhenghua An$^2$, Susumu Komiyama$^3$, Svend-Age Biehs$^{1}$}
\email{ s.age.biehs@uni-oldenburg.de} 
\affiliation{1 Institut f{\"u}r Physik, Carl von Ossietzky Universit{\"a}t, D-26111 Oldenburg, Germany}
\affiliation{2 State Key Laboratory of Surface Physics and Key Laboratory of Micro and Nano Photonics Structures (Ministry of Education), Department of Physics, Fudan University, Shanghai 200433, PR China}
\affiliation{3 Department of Basic Science, the University of Tokyo, Komaba 3-8-1, Meguro-ku, Tokyo 153-8902, Japan}



\begin{abstract}
We determine the scattered near-field and directly emitted power of a heated spherical nanoparticle above a sample within the framework of 
fluctuational electrodynamics using the dipole approximation. We find deviation of our results from previously obtained results. 
Additionally, we show that in a configuration where the nanoparticle is heated with respect to its environment the scattered power of the
near field of the sample is strictly zero. Only when the sample is heated or the temperature of the surrounding of the sample is lowered 
the scattered power will give a contribution. Our results indicate that for the interpretation of near-field imaging setups as the Thermal Infrared Near-field Spectroscope (TINS) not only the scattering of the near-field but also the direct emission of the tip plays a role.
\end{abstract}

\maketitle

%

\section{Introduction}

In the last decades different kinds of scanning thermal microscopes have been developed, which enable one to measure the thermal 
near-field either by scattering it into the far field, or by direct near-field energy transfer. A first near-field scanning thermal 
microscope of such kind has been build up in the research group of De Wilde~\cite{DeWildeEtAl2006}. This so-called  
Thermal Radiation Scanning Tunneling Microscope (TRSTM) in principle is a simple s-SNOM, which works without any external illumination. 
Therefore it scatters the thermal near-field of a heated sample at the apex of the sharp TRSTM tip into the far field. The far-field 
signal obtained in this manner can be decomposed into its different frequency parts, so that the TRSTM makes it possible to measure 
spectra of the thermal near-field in the vicinity of a sample~\cite{BabutyEtAl2013}. In order to obtain signals which are large enough 
to be measurable it is necessary to heat the samples by 100 to 200 K.  

A similar AFM-based near-field scanning thermal microscope, the so-called Thermal Infrared Near-field Spectroscope (TINS) has been set up in the group of 
Raschke~\cite{JonesEtAl2012,CallahanEtAl2014}. A key difference between the TINS and the TRSTM consists in heating up the microscope tip for TINS instead of the sample. The far-field signal emitted by the heated tip can again be decomposed in its frequency components using Fourier-Transform Infrared Spectroscopy (FTIR). Therefore, both TINS and TRSTM allow for measuring the spectra of the thermal near-field of a given sample. 

A third near-field scanning thermal microscope, the so-called Scanning Noise Microscope (SNoiM) has been constructed in the group of 
Komiyama and has been advanced by the group of An~\cite{KajiharaEtAl2010,KajiharaEtAl2011,WengEtAl2018}. As the TRSTM, the SNoiM in 
principle is an s-SNOM without external illumination. The important difference between the SNoiM and the TRSTM consists in the fact that the SNoiM has an ultra-sensitive single-photon detector which is working at cryogenic temperatures of 4.2 K. Due to this specific detector even very weak signals can be measured, to the effect that for the SNoiM measurement it is not necessary to heat up the microscope tip or the sample, as it is necessary for the TRSTM or TINS. Consequently, SNoiM is used to measure signals keeping the microscope and the sample at room temperature. However, in contrast to the TRSTM or TINS the present SNoiM setup can only measure signals at a single wavelength of 
14.5 micrometers~\cite{WengEtAl2018}. Therefore it cannot be used to get a full spectrum of the thermal near-field in its current 
stage of development.


The first theoretical models for the TRSTM, TINS, and  SNoiM were based on the assumption that the foremost part of the microscopes' 
probes can be regarded as a small sphere which can be described as dipole in the long-wavelength regime. Therefore, to lowest order the 
signals should be proportional to the photonic local density of states (LDOS) at the position of the microscope tip~\cite{DeWildeEtAl2006,JonesEtAl2012,CallahanEtAl2014,KajiharaEtAl2010,WengEtAl2018}. As a consequence, the measured signals were compared with the LDOS. An improved theoretical description has been brought forward recently by directly calculating the near-field signal which is scattered by a small sphere or ellipsoid into the far-field~\cite{JoulainEtAl2014,CvitkovicEtAl2007,JarzembskiEtAl2017}. These models seem to give quite good agreement with the measured spectra in TRSTM and TINS~\cite{BabutyEtAl2013,JarzembskiEtAl2017} 

In this work, we aim to revisit the theoretical description of the signal in the TINS based on the dipole model. In this model, one assumes
that the AFM tip can be approximated by a nanosized spherical particle which has a temperature $T_\rp$ larger than the temperature
of the surrounding $T_\rs$ consisting mainly of the sample. The power of thermal radiation $P^{\rm p}$ directly emitted by this spherical particle will be measured by the detector. Furthermore, one might expect that also the near-field of the sample which is scattered into the far field by the particle will give another contribution $P^{\rm sc}$ to the measured signal. Both contributions have been determined previously in Ref.~\cite{JoulainEtAl2014} using the concept of dressed polarizability and fluctuational electrodynamics. Here, we will show that the scattered power can only contribute to the TINS signal if the sample is heated. Additionally, we show that this scattered contribution $P^{\rm sc}$ is for distances $z_\rp \gg R$ in general much smaller than the direct emission $P^{\rm p}$ for AFM tips which are much smaller than the thermal wavelength $\lambda_{\rm th}$ which is about 10 \textmu m for temperatures around $300\,{\rm K}$. Therefore, we show that models based on the scattering approach as in Refs.~\cite{CvitkovicEtAl2007,JonesEtAl2012,CallahanEtAl2014,JarzembskiEtAl2017} are probably not appropriate for the description of the TINS signal. By comparison of our expression for $P^{\rm p}$ with the corresponding one from Ref.~\cite{JoulainEtAl2014} we find that the expression in Ref.~\cite{JoulainEtAl2014} is incorrect. Finally, we present  some numerical results for the spectral power of the scattered and direct contribution as well as the results for the full distant dependent power for both cases and discuss certain consequences for TINS.

Our work is organized as follows: In Sec.~II we introduce the dipole model and the considered configuration. The contribution of the emitted heat of a nanoparticle above a planar sample is determined in Sec.~IV based on the fluctuational fields created by the nanoparticle from Sec.~III. The resulting expression for the emitted power is compared with an existing one from the literature in Sec.~V and some simple limiting cases are discussed in Sec.~VI. In Sec.~VII we determine the scattered power, i.e.\ the contributions of the induced dipole moments due to the interaction of the nanoparticle with the fluctuating fields of the environment yielding the known expression for the scattered thermal power. The full scattered power is then determined in Sec.~VIII and numerical results are discussed in Sec.~IX and compared with experimental results from TINS. Finally, in Sec.~X we conclude with a discussion and short summary.

\section{Theoretical Model}

We want to derive an analytical expression for the amount of energy per second which is emitted by a thermally radiating 
small spherical particle with radius $R$ at a distance $z_{\rm p}$ above a planar surface of a substrate. To this end we 
assume that the particle is in local thermal equilibrum at a given temperature $T_{\rm p}$ surrounded by an environment at temperature $T_\rs$ 
and that it can be described within the dipole model. The dipole model is valid if the particle has a radius $R \ll \lambda_{\rm th}$ much 
smaller than the dominant thermal wavelength $\lambda_{\rm th}$. Furthermore, the dipole model is valid for distances $z_{\rm p}$ sufficiently larger than the radius. For distances $z_{\rm p}$ on the same order as the radius contributions of higher multipole moments have to be added to have 
a full description~\cite{KruegerEtAl2011,OteyFan2011}. 

\begin{figure}
  \epsfig{file = 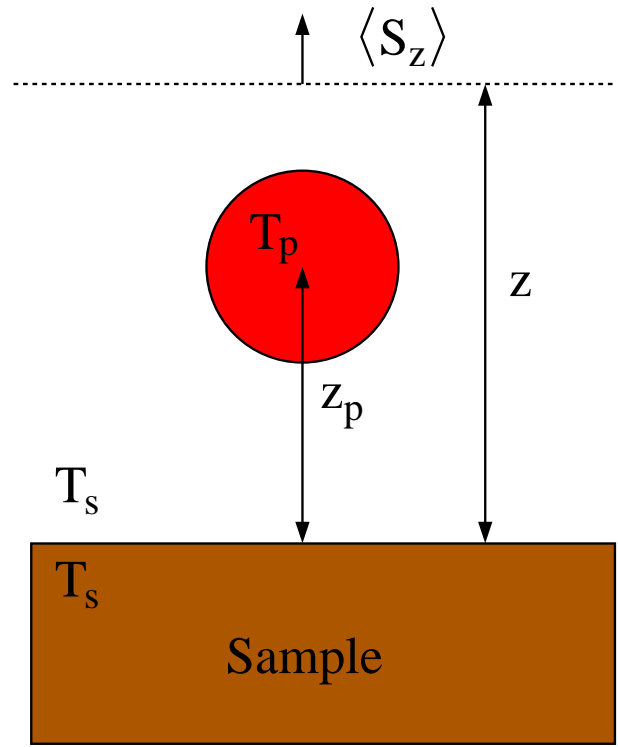, width = 0.45\textwidth}
  \caption{\label{Fig:Sketch} Sketch of the considered configuration: A nanoparticle of radius $R$ is in a distance $z_\rp$ above a sample. The surrounding environment of the nanoparticle is in thermal equilibrium at temperature $T_\rs$ and the nanoparticle is slightly heated up or cooled down so that it has the temperature $T_\rp$. The emitted power into the far-field which can be detected in the TINS setup can be determined by integrating the z component of the mean Poynting vector over the surface at position $z$.}
\end{figure}

With these asumptions the mean power emitted into the far-field is determined by the expression
\begin{equation}
  \langle P^{\rm p} \rangle = \int_{z > z_{\rm p}}\!\!\rd^2 x\, \langle S_z (z) \rangle,
\label{Eq:PowerStart}
\end{equation}
where the brackets symbolize the ensemble average. The surface integral of the z component of the mean Poynting vector 
\begin{equation}
   \langle S_z (z) \rangle = \int_0^\infty\!\!\frac{\rd \omega}{2 \pi} \, \langle S_{\omega,z} \rangle
\end{equation}
with ($j,k = x,y,z$)
\begin{equation}
   \langle S_{\omega,z} \rangle = 2 \Re \bigl(\epsilon_{zjk} \langle E_j(\mathbf{r},\omega) H_k^*(\mathbf{r},\omega) \rangle \bigr)
\end{equation}
is performed over a planar surface parallel to the surface of the substrate at a distance $z$ greater than $z_{\rm p}$ as 
sketched in Fig.~\ref{Fig:Sketch}. Note that here we use the Einstein convention and we have introduced the Levi-Civita tensor $\epsilon_{ijk}$. 

\section{Fluctuating Fields}

In order to determine the mean Poynting vector it is necessary to calculate the fluctuating electromagnetic fields
$\mathbf{E}(\mathbf{r},\omega)$ and $\mathbf{H}(\mathbf{r},\omega)$ which are produced by the thermally fluctuating
charges inside the nanoparticle. Within the dipole model these sources of the fluctuating fields can be described by
a fluctuating dipole moment $\mathbf{p}^{\rm fl}$ at the position $\mathbf{r}_{\rm p}$ of the nanoparticle. The total 
electric field at position $\mathbf{r}$ can then be expressed formally as
\begin{equation}
  \mathbf{E}(\mathbf{r},\omega) = \mu_0 \omega^2 \mathds{G}^{\rm E}(\mathbf{r,r_{\rp}},\omega)\cdot \mathbf{p}
\label{Eq:field}
\end{equation}
introducing the electric Green function $\mathds{G}^{\rm E}(\mathbf{r,r_{\rp}})$ of the given environment and the total dipole moment 
\begin{equation}
  \mathbf{p} =  \mathbf{p}^{\rm fl} + \mathbf{p}^{\rm ind}
\end{equation}
which consists of the fluctuating dipole moment $\mathbf{p}^{\rm fl}$ describing the thermal sources inside the nanoparticle and the
induced dipole moment $\mathbf{p}^{\rm ind}$. For an homogeneous isotropic and spherical nanoparticle with permittivity 
$\epsilon_{\rm p} (\omega)$ the relation between the induced dipole moment and the electric field is
\begin{equation}
   \mathbf{p}^{\rm ind} = \epsilon_0 \tilde{\alpha}(\omega) \mathbf{E} (\mathbf{r}_{\rm p},\omega) 
\end{equation}
with the 'naked' polarizability
\begin{equation}
   \tilde{\alpha} (\omega) = 4 \pi R^3 \frac{\epsilon_{\rm p} (\omega) - 1}{\epsilon_{\rm p} (\omega) + 2}.
\end{equation} 
Inserting the expression for the total dipole moment into Eq.~(\ref{Eq:field}) we find 
\begin{equation}
  \mathbf{E}(\mathbf{r},\omega) = \mu_0 \omega^2 \mathds{G}^{\rm E}(\mathbf{r,r_{\rp}},\omega)\cdot \tilde{\mathds{D}}^{-1} \cdot \mathbf{p}^{\rm fl}
\label{Eq:Efieldfinal}
\end{equation}
which is the expression of the full fluctuating electric field produced by the fluctuating sources
taking into account all the multiple interactions with the environment via the term
\begin{equation}
   \tilde{\mathds{D}} = \mathds{1} - k_0^2 \tilde{\alpha} \mathds{G}^{\rm E}(\mathbf{r_{\rp},r_{\rp}},\omega)
\label{Eq:Dtilde}
\end{equation}
where $\mathds{1}$ is the unit matrix and $k_0 := \omega/c$ is the wave number in free space. Since the electric and magnetic field
are connected by Faraday's law
\begin{equation}
  \mathbf{H}(\mathbf{r},\omega) = \frac{1}{\ri \omega \mu_0} \nabla\times\mathbf{E}(\mathbf{r},\omega)
\end{equation} 
we can instantly express the magnetic fluctuating field produced by the thermal sources as
\begin{equation}
   \mathbf{H}(\mathbf{r},\omega) = \mu_0 \omega^2 \mathds{G}^{\rm H}(\mathbf{r,r_{\rp}},\omega)\cdot \tilde{\mathds{D}}^{-1} \cdot \mathbf{p}^{\rm fl}
\label{Eq:Hfieldfinal}
\end{equation}
by introducing the magnetic Green's function as
\begin{equation}
  \mathds{G}^{\rm H}(\mathbf{r,r_{\rp}},\omega) =  \frac{1}{\ri \omega \mu_0} \nabla\times \mathds{G}^{\rm E}(\mathbf{r,r_{\rp}},\omega).
\end{equation}

Now, the Green tensor can be decomposed in a vacuum part and a scattered part
\begin{equation}
  \mathds{G}^{\rm E}(\mathbf{r_{\rp},r_{\rp}},\omega) = \mathds{G}^{\rm E,0} (\mathbf{r_{\rp},r_{\rp}},\omega) + \mathds{G}^{\rm E,sc} (\mathbf{r_{\rp},r_{\rp}},\omega).
\end{equation}
It is well-known that the real part of the Green tensor $\mathds{G}^{\rm E,0}(\mathbf{r_{\rp},r_{\rp}},\omega)$ is divergent so that 
$\mathds{D}$ is divergent as well. This unphysical results stems from the fact that the nanoparticle has been replaced by a point dipole.
As described in large detail in Refs.~\cite{AlbadalejoEtAl2010} this anomaly can be circumvented by considering a finite particle. In this case one obtains
similar results as above but with $\tilde{\mathds{D}}$ containing only the scattered part of the Green's function and the polarizability
$\tilde{\alpha}$ replaced by the dressed polarizability $\alpha$. Hence, the correct expressions for the fluctuating fields are
\begin{align}
   \mathbf{E}(\mathbf{r},\omega) &= \mu_0 \omega^2 \mathds{G}^{\rm E}(\mathbf{r,r_{\rp}},\omega)\cdot{\mathds{D}}^{-1} \cdot \mathbf{p}^{\rm fl}, \\
   \mathbf{H}(\mathbf{r},\omega) &= \mu_0 \omega^2 \mathds{G}^{\rm H}(\mathbf{r,r_{\rp}},\omega)\cdot {\mathds{D}}^{-1} \cdot \mathbf{p}^{\rm fl}
\end{align}
with
\begin{equation}
   {\mathds{D}} = \mathds{1} - k_0^2 {\alpha} \mathds{G}^{\rm E,sc}(\mathbf{r_{\rp},r_{\rp}},\omega)
\end{equation}
and the dressed polarizability
\begin{equation}
  \alpha(\omega) = \frac{\tilde{\alpha}(\omega)}{1 - \ri \frac{k_0^3}{6 \pi} \tilde{\alpha}(\omega)}.
\end{equation}
Note, that for the isotropic nanoparticle the matrix $\mathds{D}$ is diagonal so that it can be written in 
the simple form $\mathds{D}_{ij} = D_{ii} \delta_{ij}$ with the diagonal elements $D_{ii}$.
{It is known that the radiation correction term $-\ri \frac{k_0^3}{6 \pi} \tilde{\alpha}(\omega)$ in the dressed polarizability is important for energy conservation~\cite{YurkinHoekstra2007}. Here, in the context of thermal radiation, it will be important to have a vanishing heat flux at thermal equilibrium as will be seen in Sec.~VIII.}

\section{Emitted Power}

With the expressions for the fields in terms of the Green function we can now evaluate the z component of the mean spectral Poynting vector.
First, we obtain straight forwardly
\begin{equation}
\begin{split}
  \langle S_{\omega,z}\rangle & =  \varepsilon_{zjk} \langle E_j H_k^* \rangle + \mathrm{c.c.} \\
                            & =  \mu_0^2 \omega^4 \varepsilon_{zjk} \mathds{G}_{jl}^{\mathrm{E}} \mathds{D}_{ll}^{-1} \mathds{G}_{km}^{\mathrm{H} *} \mathds{D}_{mm}^{-1 *} \langle p_l^{\mathrm{fl}} p_m^{\mathrm{fl}^*} \rangle +\mathrm{c.c.}
\label{eq:S_1}
\end{split}
\end{equation}
This expression can be further simplified by making use of the fluctuation-dissipation theorem for the fluctuating
dipole moments which is given by~\cite{MessinaEtAl2013}
\begin{equation}
   \langle p_l^{\mathrm{fl}} p_m^{\mathrm{fl}^*} \rangle = \frac{2 \varepsilon_0 \chi}{\omega} \Theta(T_\rp) \delta_{lm}
\end{equation}
with 
\begin{equation}
   \chi = \Im(\alpha) - \frac{k_0^3}{6 \pi} |\alpha|^2 
             = \frac{\Im(\tilde{\alpha})}{\bigl|1 - \frac{\ri k_0^3}{6 \pi} \tilde{\alpha}\bigr|^2}  
\label{Eq:Chi}
\end{equation}
and the thermal part of the mean energy of a harmonic oscillator
\begin{equation}
  \Theta(T) = \frac{\hbar \omega}{e^{\frac{\hbar \omega}{k_B T}} - 1} .
\label{eq:Planck}
\end{equation}
Inserting the fluctuation-dissipation theorem in the expression for the mean spectral Poynting vector we find
\begin{equation}
  \langle S_{\omega,z}\rangle = 2 \mu_0 \frac{\omega^3}{c^2} \chi \Theta(T_\rp) \varepsilon_{zjk} \mathds{G}_{jl}^{\mathrm{E}} \bigl|\mathds{D}_{ll}\bigr|^{-2} \mathds{G}_{lk}^{\mathrm{H}^\dagger} +\mathrm{c.c.}
\label{Eq:PoyntingZw}
\end{equation}

Before we can further evaluate the Poynting vector we need to specify the electric and magnetic Green function for our given geometry. It can be expressed as
\begin{equation}
   {\mathds{G}}^{\mathrm{E/H}}(\boldsymbol{r},\boldsymbol{r}_p,\omega) =  \int \frac{\mathrm{d}^2\kappa}{(2\pi)^2}e^{\mathrm{i} \boldsymbol{\kappa}\cdot(\boldsymbol{x}-\boldsymbol{x_p})}{\mathds{G}}^{\mathrm{E/H}}(\boldsymbol{\kappa},z,\omega)
\label{eq:Green}
\end{equation}
using the notation  $\mathbf{x} = (x , y)^t$ and $\kappa = (k_x, k_y)^t$ with the integrand
\begin{equation}
\begin{split}
  {\mathds{G}}^{\mathrm{E}}(\boldsymbol{\kappa},z,\omega) &= \frac{\mathrm{i}e^{\mathrm{i} \gamma_0(z-z_p)}}{2\gamma_0}\left[\sum_{i = \rs,\rp}\boldsymbol{a}_i^+(k_{0})\otimes\boldsymbol{a}_i^+(k_{0})\right] \\
     &+ \frac{\mathrm{i}e^{\mathrm{i} \gamma_0(z+z_p)}}{2\gamma_0}\left[\sum_{i = \rs,\rp} r_i\boldsymbol{a}_i^+(k_0)\otimes\boldsymbol{a}_i^-(k_0) \right].
\end{split}
\label{eq:Green_kappa}
\end{equation}
The magnetic expression ${\mathds{G}}^{\mathrm{H}}(\boldsymbol{\kappa},z,\omega)$ can be obtained from ${\mathds{G}}^{\mathrm{E}}(\boldsymbol{\kappa},z,\omega)$ by simply replacing the left hand vector $\mathbf{a}_\rp^+$ by $\mathbf{a}_\rs^+/\omega\mu_0$ and $\mathbf{a}_\rs^+$ by $-\mathbf{a}_\rp^+/\omega\mu_0$. These vectors are the polarization vectors for the s- and p-polarization\begin{align}
     \boldsymbol{a}_s^\pm(k) &= \frac{\boldsymbol{k}_\pm \times \boldsymbol{e}_z}{\sqrt{\left(\boldsymbol{k}_\pm \times \boldsymbol{e}_z\right)^2}}, 
  \label{eq:a_s}\\
     \boldsymbol{a}_p^\pm(k) &= \frac{\boldsymbol{a}_s^\pm \times \boldsymbol{k}_\pm}{\sqrt{\left(\boldsymbol{a}_s^\pm \times \boldsymbol{k}_\pm\right)^2}} 
  \label{eq:a_p}
\end{align}
which are defined with respect to the surface normal $\mathbf{e}_z$ and the wave vector $\mathbf{k}_\pm = (\kappa, \pm \gamma_0)$, where we
have introduced the wavenumber in vacuum perpendicular to the interface $\gamma_0 = \sqrt{k_0^2 - \kappa^2}$. Furthermore, we have introduced the Fresnel reflection coefficients for s- and p-polarized light
\begin{align}
   r_s & = \frac{\gamma_0 - \gamma_1}{\gamma_0 + \gamma_1}, \label{eq:r_s} \\
   r_p & = \frac{\varepsilon_1 \gamma_0 - \gamma_1}{\varepsilon_1 \gamma_0 + \gamma_1}
\label{eq:r_p}
\end{align}
which contain the permittivity of the substrate $\epsilon_1 (\omega)$ and the wavenumber perpendicular to the interface inside of the medium $\gamma_1 = \sqrt{k_0^2 \epsilon_1 - \kappa^2}$.

Now, we have finally all the ingredients to determine the emitted power in Eq.~(\ref{Eq:PowerStart}). Eventually we obtain the final expression for the spectral emitted power which can be written as
\begin{equation}
        P_\omega^{\rm p} = k_0^3 \chi \Theta(T_\rp) \biggl[ \frac{I_{xx}}{|D_{xx}|^2} + \frac{I_{zz}}{ |D_{zz}|^{2}}\biggr] \mathrm{.}
\label{eq:P}
\end{equation}
with 
\begin{align}
   \begin{split}
   I_{xx} &= \int_0^{k_0} \!\! \frac{\mathrm{d} \kappa}{2 \pi} \, \frac{\kappa}{k_0 \gamma_0}\biggl[\bigl|1 + r_s e^{2 \mathrm{i} \gamma_0 z_p} \bigr|^2 \\
          &\qquad + \left(1 - \frac{\kappa^2}{k_0^2}\right) \bigl| 1 - r_p  e^{2 \mathrm{i} \gamma_0 z_p}  \bigr|^2 \biggr], 
   \end{split}\\
   I_{zz} &=  \int_0^{k_0} \!\! \frac{\mathrm{d} \kappa}{2 \pi} \, \frac{\kappa^3}{k_0^3 \gamma_0} \bigl|1 + r_p e^{2 \mathrm{i} \gamma_0 z_p} \bigr|^2 \biggr]
\end{align}
and
\begin{equation}
  D_{xx/zz}  = 1 - k_0^2 \alpha {\mathds{G}}_{xx/zz}^{\mathrm{EE,Sc}}(\mathbf{r}_\rp,\mathbf{r}_\rp) \mathrm{.}
\label{eq:d_xx_-1} 
\end{equation}
As expected, the emitted power does not depend on the position $z$ of the surface over which we have integrated and
it also does not depend on the lateral position $\mathbf{x}_p = (x_p,y_p)^t$ of the particle due to the translation symmetry
within the x-y plane. But of course, the emitted power depends on the distance $z_p$ of the particle to the interface. Note, that
the emitted heat flux is due to propagating waves with $\kappa \leq k_0$ only. Nonetheless, evanescent modes with $\kappa > k_0$
contribute in the near-field interaction with the surface via the terms~\cite{Doro2011}
\begin{align}
 \!\!\!\!\!\!\! {\mathds{G}}_{xx}^{\mathrm{EE,sc}}(\mathbf{r}_\rp,\mathbf{r}_\rp) &= \!\!\!  \int_0^{\infty} \!\!  \frac{\mathrm{d}\kappa}{2 \pi} \, \frac{\ri \kappa \re^{2 \ri \gamma_0 z_\rp }}{4 \gamma_0} \biggl( r_\rs - r_\rp \frac{\gamma_0^2}{k_0^2} \biggr),  \\
   {\mathds{G}}_{zz}^{\mathrm{EE,sc}}(\mathbf{r}_\rp,\mathbf{r}_\rp) &= \!\!\! \int_0^{\infty} \!\!  \frac{\mathrm{d}\kappa}{2 \pi} \, \frac{\ri \kappa \re^{2 \ri \gamma_0 z_\rp }}{4 \gamma_0} 2 r_\rp \frac{\kappa^2}{k_0^2} 
\end{align}
in the prefactors $D_{xx/zz}$. 

Before we compare our result to the literature, let us rewrite the expression for $P(\omega)$ in terms of the undressed polarizability $\tilde{\alpha}$. To this end, we insert Eq.~(\ref{Eq:Chi}) into Eq.~(\ref{eq:P}) and obtain
\begin{equation}
        P_\omega^{\rm p} = k_0^3 \Im(\tilde{\alpha}) \Theta(T_\rp) \biggl[ \frac{I_{xx}}{|\tilde{D}_{xx}|^{2}}  + \frac{I_{zz}}{|\tilde{D}_{zz}|^{2}} \biggr].
\label{eq:Pnaked}
\end{equation}
with 
\begin{equation}
  \tilde{D}_{xx/zz}  = 1 - \frac{\ri k_0^3}{6 \pi} \tilde{\alpha} - k_0^2 \tilde{\alpha} {\mathds{G}}_{xx/zz}^{\mathrm{EE,Sc}}(\mathbf{r}_\rp,\mathbf{r}_\rp) \mathrm{.}
\label{eq:d_xx_-1naked} 
\end{equation}
From the expressions for the denominators $\tilde{D}_{xx/zz}$ it becomes obvious that the above procedure for regulating the divergence of $\mathds{G}^{E,0}(\mathbf{r}_\rp,\mathbf{r}_\rp)$ by introducing the dressed polarizability is equivalent to replacing the divergent term $- k_0^2 \tilde{\alpha}\mathds{G}^{E,0}(\mathbf{r}_\rp,\mathbf{r}_\rp)$ in the expression of $\tilde{\mathds{D}}$ in Eq.~(\ref{Eq:Dtilde}) by the radiation correction $- \frac{\ri k_0^3}{6 \pi} \tilde{\alpha}$. 

\section{Comparison with literature}

First, we want to compare our result with an expression for the heat flux which has been given in Eq.~(44) in Ref.~\cite{JoulainEtAl2014}. There the expression
\begin{equation}
        P_\omega'  = k_0^3  \Theta(T_\rp) \biggl[ \Im(\alpha_{xx}) I_{xx} + \Im(\alpha_{zz}) I_{zz} \biggr]
\label{eq:Pprime}
\end{equation}
with 
\begin{equation}
  \alpha_{xx/zz}  = \frac{\tilde{\alpha}}{1 - k_0^2 \tilde{\alpha} {\mathds{G}}_{xx/zz}^{\mathrm{EE,Sc}}(\mathbf{r}_\rp,\mathbf{r}_\rp)} \mathrm{.}
\label{eq:alphaxx} 
\end{equation}
 is given without derivation. We have simplified the expression for the the electric response and present it here in our notation. To compare this expression with, for example, our Eq.~(\ref{eq:Pnaked}) we explicit the formulas for $\Im(\alpha_{xx/zz})$ which are
\begin{equation}
  \Im(\alpha_{xx/yy}) = \frac{\Im(\tilde{\alpha}) + |\tilde{\alpha}|^2 k_0^2 \Im(\mathds{G}^{E,sc}_{xx/zz})}{|D_{xx/zz}|^2}.
\end{equation}
This expression is obviously different from ours. In particular, it does not contain the radiation losses in the denominator. These radiation losses are in the infrared for nanoparticles in most cases negligble so that this difference is not very crucial. What is more disturbing is the fact that the expression for $P'_\omega$ gives by virtue of the second term in $\Im(\alpha_{xx/yy})$ for the flux a contribution proportional to $\Im(\mathds{G}^{E,sc}_{xx/zz})$, i.e.\ an emitted power which is proportional to mainly the partial photonic electrical local density of states~\cite{Agarwal1975, Eckardt, JoulainSurfSciRep05}. {red}{Note that there should be no direct evanescent contribution to the radiated heat flux as suggested by the expression $P_\omega'$. The heat flux is entirely due to propagating waves. Hence, our expression $P_\omega^{\rm p}$ is also a qualitative correction of the expression $P_\omega$ even though the quantitative difference is in most cases small.} Since in Ref.~\cite{JoulainEtAl2014} there is no derivation of the expression (\ref{eq:Pprime}) we have no possibility to trace back the source of the difference. Nonetheless, we will see in Sec.~VIII that a part of the scattered power has the form of Eq.~(\ref{eq:Pprime}) but with temperature $T_\rs$ instead of temperature $T_\rp$.

\section{Limiting cases}

Let us first consider the case of a perfect metal with  $r_\rs = - 1$ and  $r_\rp = 1$ taking $z_\rp \rightarrow \infty$. Then we obtain
\begin{equation}
        P_\omega^{\rm p} \rightarrow  2 \Theta(T_\rp) \frac{k_0^3}{\pi} \frac{\Im(\tilde{\alpha})}{\bigl|  1 - \frac{\ri k_0^3}{6 \pi} \tilde{\alpha} \bigr|^2} \equiv P^{\rm vac}(\omega). 
\end{equation}
This result should correspond to the full power $P^{\rm vac}(\omega)$ emitted by a nanoparticle in free space, because due to the totally reflecting substrate all the energy which is emitted into negative z direction will be reflected. By taking $z_\rp \rightarrow \infty$ we have furthermore turned off possible interference effects of the emitted and reflected thermal light. By comparing $P^{\rm vac}(\omega)$ with the textbook value in Ref.~\cite{BohrenHuffman} we find full agreement if we neglect the radiation correction term $-\frac{\ri k_0^3}{6 \pi} \tilde{\alpha}$ which has not been considered in Ref.~\cite{BohrenHuffman}.  Now, by setting the reflection coefficients $r_\rs = r_\rp = 0$ we obtain from Eq.~(\ref{eq:Pnaked})
\begin{equation}
        P_\omega^{\rm p}  = \frac{1}{2} P^{\rm vac}(\omega). 
\label{Eq:PparticleBB}
\end{equation}
Therefore we obtain merely half of the emitted power $P^{\rm vac}(\omega)$ because our expression only takes into account the part emitted in positive z direction. We note, that the erroneous expression $P'(\omega)$ in Eq.~(\ref{eq:Pprime}) gives the correct limits for both cases but without the radiation correction.

For very small distances the emitted power is given by a constant value which is determined by the expression for $P(\omega)$ setting $z_\rp = R$. For large distances $z_\rp \rightarrow \infty$ it is easy to show that the emitted power converges to the distance-independent value
\begin{equation}
   P_\omega^{\rm p} \rightarrow \frac{P^{\rm vac}}{4} \left[2 + \int_0^{k_0} \!\!\! \mathrm{d} \kappa \, \frac{\kappa}{k_0 \gamma_0}\left(|r_s|^2 + |r_p|^2 \right) \right] \equiv P^\infty(\omega).
\label{eq:P_unendlich}
\end{equation}
From this expression one can instantly see that the emitted power in the case of $r_\rp = r_\rs = 0$ results in $P^{\rm vac}(\omega)/2$ which is the smallest possible value for $z_\rp \rightarrow \infty$. On the other hand for the perfect metal case one obtains again $P^{\rm vac}(\omega)$. Therefore we have $P^{\rm vac}/2 \leq P \leq P^{\rm vac}$ for $z_\rp \rightarrow \infty$. For intermediate distances we can employ the method of stationary phase to obtain a good approximation of the emitted power. We find
\begin{equation}
\begin{split}
  P_\omega^{\rm p} &\approx P^\infty(\omega) \\
            &\quad+ \frac{P^{\rm vac}}{4} \frac{2}{k_0^2 z_p} \mathrm{Re} \biggl(\frac{1 - \sqrt{\varepsilon(\omega)}}{1 + \sqrt{\varepsilon(\omega)}} e^{\mathrm{i} (2 k_0 z_p - \frac{\pi}{4})}\biggr). 
\end{split}
\label{Eq:ApproxPdPerfectMetal}
\end{equation}
Obviously, the emitted power is oscillating around the value $P^\infty(\omega)$ due to the interference of the emitted power in the positive z direction and the reflected power. It is expected that this interference effect sets in for distances $z_\rp$ on the order of the wavelength. For the integrated power we expect therefore an oscillatory behaviour for distances on the order of $\lambda_{\rm th}$.

\section{Scattered Power}

Now, we want to contrast the results for the emitted power of the nanoparticle with the scattered power. This can be done, by assuming that
the particle is immersed in the thermally fluctuating electric and magnetic fields $\mathbf{E}^\rs$ and $\mathbf{H}^\rs$ of its surrounding which
is considered to be in equilibrium at a temperature $T_\rs$. In this case the total field at a position $\mathbf{r}$ can be described by the
fields of the surrounding plus the contribution of the induced dipole moment of the nanoparticle
\begin{equation}
\begin{split}
  \mathbf{E}(\mathbf{r}) &= \mu_0 \omega^2 \mathds{G}^{\rm E}(\mathbf{r},\mathbf{r}_\rp)\cdot\mathbf{p}^{\rm ind} + \mathbf{E}^\rs(\mathbf{r}) \\
                         &= k_0^2 \tilde{\alpha} \mathds{G}^{\rm E}(\mathbf{r},\mathbf{r}_\rp)\cdot\mathbf{E}(\mathbf{r}_\rp) + \mathbf{E}^\rs(\mathbf{r}).
\end{split}
\end{equation}
Solving this equation for the total field at the position of the particle gives $\mathbf{E}(\mathbf{r}_\rp) = \tilde{\mathds{D}}^{-1}\cdot\mathbf{E}^\rs(\mathbf{r}_\rp)$. Hence we encounter again the problem of the divergence in $\tilde{\mathds{D}}$. With the same procedure as before we find 
\begin{equation}
   \mathbf{E}(\mathbf{r}) = k_0^2 \alpha \mathds{G}^{\rm E}(\mathbf{r},\mathbf{r}_\rp)\cdot\mathds{D}^{-1}\cdot\mathbf{E}^{\rs}(\mathbf{r}_\rp) + \mathbf{E}^\rs(\mathbf{r})
\label{Eq:FieldscatteredFull}
\end{equation}
using the dressed polarizability. The first term clearly gives the contribution of the induced dipole moment, so that the electric and magnetic fields scattered by the nanoparticle are given by
\begin{align}
   \mathbf{E}^{\rm sc}(\mathbf{r},\omega) &= k_0^2 \alpha \mathds{G}^{\rm E}(\mathbf{r,r_{\rp}},\omega)\cdot{\mathds{D}}^{-1} \cdot \mathbf{E}^{\rs}(\mathbf{r}_\rp), \\
   \mathbf{H}^{\rm sc}(\mathbf{r},\omega) &= k_0^2 \alpha \mathds{G}^{\rm H}(\mathbf{r,r_{\rp}},\omega)\cdot {\mathds{D}}^{-1} \cdot \mathbf{E}^{\rs}(\mathbf{r}_\rp).
\end{align}
Using these expressions we can determine the z component of the mean Poynting vector by virtue of the known relation~\cite{Doro2011} $\langle E_{n}^s {E_{l}^s}^* \rangle = \langle |E_l^s|^2 \rangle \delta_{nl}$ giving
\begin{equation}
  \langle S_{\omega,z}^{\rm sc}\rangle = k_0^4 |\alpha|^2 \varepsilon_{zjk} \mathds{G}_{jl}^{\mathrm{E}} \bigl|\mathds{D}_{ll}\bigr|^{-2} \mathds{G}_{lk}^{\mathrm{H}^\dagger} \langle |E_l^\rs|^2 \rangle +\mathrm{c.c.}
\end{equation}
This Poynting vector has clearly exactly the same mathematical structure as the expression in Eq.~(\ref{Eq:PoyntingZw}). Obviously, both equations can be converted into one another by making the replacement
\begin{equation}
  \frac{2 \mu_0 \omega^3}{c^2} \chi \Theta(T_\rp) |D_{ll}|^{-1} \quad \leftrightarrow \quad |\alpha|^2 k_0^4 |D_{ll}|^{-1} \langle |E_l^\rs|^2 \rangle.
\end{equation}
Therefore we can directly obtain from the expression in Eq.~(\ref{eq:Pnaked}) the power scattered in positive z direction. We obtain for 
the scattered power
\begin{equation}
        P^{\rm sc}_\omega = c k_0^4 |\tilde{\alpha}|^2  \biggl[ \frac{\frac{\epsilon_0}{2}\langle |E_x^\rs|^2 \rangle}{|\tilde{D}_{xx}|^{2}} I_{xx} + \frac{\frac{\epsilon_0}{2}\langle |E_z^\rs|^2 \rangle}{|\tilde{D}_{zz}|^{2}} I_{zz}  \biggr] \mathrm{.}
\label{eq:Pscat}
\end{equation}
This expression is apart from the  radiation correction the same expression as in Eq.~(38) from Ref.~\cite{JoulainEtAl2014} and it is the same as the corresponding expression in~\cite{KallelEtAl2017}. To evaluate it we need the
well-known expressions for the x and z components of the mean spectral electrical energy density which is given as~\cite{Eckardt, JoulainSurfSciRep05,Doro2011}
\begin{equation}
\begin{split}
   \frac{\epsilon_0}{2}\langle |E_{x/z}^\rs|^2 \rangle &= \frac{k_0}{c} \Theta(T_\rs) \Im \mathds{G}^{\rm E}_{xx/zz} \\
                                                       &= \frac{k_0}{c} \Theta(T_\rs) \biggl( \frac{k_0}{6 \pi} + \Im \mathds{G}^{\rm E,sc}_{xx/zz}  \biggr).
\end{split}
\end{equation}
As for the radiated power the scattered power is due to the propagating modes with $\kappa \leq k_0$. Nonetheless, due to the interaction with the near field of the interface there is also an evanescent contribution in $\langle |E_{x/z}^\rs|^2 \rangle$ and $\tilde{D}_{xx/zz}$. It has to be noted that for $z_p \rightarrow 0$ the scattered power does not diverge because $\langle |E_{x/z}^\rs|^2 \rangle \propto 1/z_\rp^3$ but $|\tilde{D}_{xx/zz}|^2 \propto 1/z_\rp^6$! {This is due to the term proportional to the scattered part of the Green function in $\tilde{D}_{xx/zz}$ which is very important in the near-field regime close to a resonance of the particle or surface. This is, in particular, the case in TINS where the spectra of surface mode resonances are measured. The radiation-correction term is in this situation negligibly small.} 

In a similar fashion as before we can again discuss different limiting cases. For convenience we only discuss the case $r_\rs = r_\rp = 0$. Then we simply obtain for the scattered thermal radiation the value for the scattered thermal photon gas, which is
\begin{equation}
  P^{\rm sc}_\omega = \frac{k_0^6}{6 \pi^2} \Theta(T_\rs) \frac{|\tilde{\alpha}|^2}{\bigl|  1 - \frac{\ri k_0^3}{6 \pi} \tilde{\alpha} \bigr|^2}.
\label{Eq:PscatBB}
\end{equation}  
For perfect reflection and $z_\rp \rightarrow \infty$ we obtain again twice this result. In both special cases it can be nicely seen that $P^{\rm sc}_\omega \propto (R k_0)^6$, whereas for the directly emitted part we have $P^{\rm p}_\omega \propto (R k_0)^3$. Hence one can expect to have a domination of $P^{\rm p}$ over $P^{\rm sc}$ for spheres with $R \ll \lambda$. Note, that this condition is not very strict, because close to a resonance $|\tilde{\alpha}|^2 \gg \Im(\tilde{\alpha})$.

\section{Contribution of scattered Power in TINS?} 

In the last section, we have determined the scattered power by focusing on the fields which are inducing a dipole moment in the nanoparticle. But this is not giving the full expression for the scattered power. In order to get the full expression it is necessary to determine the mean Poynting vector starting with the full expressions of the fields in Eq.~(\ref{Eq:FieldscatteredFull}), i.e.\ we need work with
\begin{align}
   \mathbf{E}(\mathbf{r}) &= k_0^2 \alpha \mathds{G}^{\rm E}(\mathbf{r},\mathbf{r}_\rp)\cdot\mathds{D}^{-1}\cdot\mathbf{E}^{\rs}(\mathbf{r}_\rp) + \mathbf{E}^\rs(\mathbf{r}), \\
   \mathbf{H}(\mathbf{r}) &= k_0^2 \alpha \mathds{G}^{\rm H}(\mathbf{r},\mathbf{r}_\rp)\cdot\mathds{D}^{-1}\cdot\mathbf{E}^{\rs}(\mathbf{r}_\rp) + \mathbf{H}^\rs(\mathbf{r}).
\end{align}
Then we obtain for the z component of the full mean Poynting vector
\begin{equation}
\begin{split}
  \langle S_{z,\omega}^{\rm sc,f} \rangle &= \langle S_{z,\omega}^{\rm sc} \rangle \\ 
                                  &\quad + \epsilon_{ijz} k_0^2 \alpha \mathds{G}^{\rm E}_{ik} D_{kk}^{-1} \langle E_k^\rs(\mathbf{r}_\rp) {H_j^\rs}^*(\mathbf{r})\rangle\\
                                  &\quad + \epsilon_{ijz} k_0^2 \alpha^* {\mathds{G}^{\rm H}_{jk}}^* {D_{kk}^{-1}}^* \langle E_i^\rs(\mathbf{r}) {E_k^\rs}^*(\mathbf{r}_\rp)\rangle\\
                                  &\quad +  \epsilon_{ijz} \langle E_i^\rs(\mathbf{r}) {H_j^\rs}^*(\mathbf{r})\rangle \\
                                  &\quad + {\rm c.c.}
\end{split}
\label{Eq:Szw}
\end{equation}
The first term on the right hand side is just the contribution discussed in the last section. One can see that there are extra terms which have to be considered in general as well. 

If we assume, as before, that the fields $\mathbf{E}^\rs$ and $\mathbf{H}^\rs$ are in thermal equilibrium at temperature $T_\rs$, then the correlation function of the electric and magnetic fields are given by~\cite{Eckhardt1984}
\begin{align}
   \langle  {H_j^\rs}^*(\mathbf{r}) E_k^\rs(\mathbf{r}_p) \rangle &= 2 \omega \mu_0 \Theta(T_\rs) \ri \Re \bigl[ \mathds{G}_{jk}^{\rm H}(\mathbf{r},\mathbf{r}_\rp)\bigr] \\
   \langle  {E_i^\rs}(\mathbf{r}) {E_k^\rs}^*(\mathbf{r}_p) \rangle &= 2 \omega \mu_0 \Theta(T_\rs) \Im \bigl[ \mathds{G}_{ik}^{\rm E}(\mathbf{r},\mathbf{r}_\rp)\bigr]
\end{align}
where we have as before neglected the zero-point contribution. Inserting the mixed correlation function of the electric and magnetic field the last term in Eq.~(\ref{Eq:Szw}) gives zero so that in our equilibrium situation the direct mean energy flow from the sourrounding field vanishes as it should be. The second and third term of Eq.~(\ref{Eq:Szw}) can be summed up so that we obtain by using the correlation functions the expression
\begin{equation}
  \langle S_{z,\omega}^{\rm sc,f} \rangle = \langle S_{z,\omega}^{\rm sc} \rangle + \langle S_{z,\omega}^{\rm extra} \rangle
\end{equation}
with
\begin{equation}
\begin{split}
  \langle S_{z,\omega}^{\rm extra} \rangle  &= - 2 \mu_0 \frac{\omega^3}{c^2} \Theta(T_\rs) \epsilon_{ijz} \Im\biggl(\frac{\alpha}{D_{kk}}\biggr) \\
                                            &\qquad\times\mathds{G}^{\rm E}_{ik}(\mathbf{r},\mathbf{r}_\rp) {\mathds{G}^{\rm H}_{jk}}^*(\mathbf{r},\mathbf{r}_\rp) + \text{c.c.}
\end{split}
\end{equation}
The extra term has again a similar structure as expresssion (\ref{Eq:PoyntingZw}) so that we can easily obtain the final result for the power flowing through the interface at position $z$ by replacing
\begin{equation}
  \chi |D^{-1}_{kk}|^2 \Theta(T_\rp)  \quad \leftrightarrow - \Im\bigl(\alpha D^{-1}_{kk}\bigr) \Theta(T_\rs).
\end{equation}
We then find that the extra term gives a spectral power of
\begin{equation}
\begin{split}
  P^{\rm extra}_\omega &= - k_0^3 \Theta(T_\rs) \biggl[ \Im\biggl(\frac{\alpha}{D_{xx}}\biggr) I_{xx} + \Im\biggl(\frac{\alpha}{D_{zz}}\biggr) I_{zz} \biggr] \\
                       &= - k_0^3 \Theta(T_\rs) \Im(\tilde{\alpha}) \biggl[ \frac{I_{xx}}{|\tilde{D}_{xx}|^2} +   \frac{I_{zz}}{|\tilde{D}_{zz}|^2} \biggr] \\
                       &\quad  - k_0^5 \Theta(T_\rs) |\tilde{\alpha}|^2  \biggl[ \frac{I_{xx} \Im \mathds{G}^{\rm E,sc}_{xx}  }{|\tilde{D}_{xx}|^2} +   \frac{I_{zz} \Im \mathds{G}^{\rm E,sc}_{zz} }{|\tilde{D}_{zz}|^2} \biggr] \\
                       &\quad - \frac{k_0^6}{6 \pi} \Theta(T_\rs) |\tilde{\alpha}|^2  \biggl[ \frac{I_{xx}}{|\tilde{D}_{xx}|^2} +   \frac{I_{zz}}{|\tilde{D}_{zz}|^2} \biggr].
\end{split}
\end{equation}
Apart from the overall sign and the  radiation correction (in $\tilde{D}_{xx/zz}$ and the last term) and the different temperature this is just the same expression as in Eq.~(\ref{eq:Pprime}). We can only guess that the expression Eq.~(\ref{eq:Pprime}) has been obtained in the same way as $P^{\rm extra}_\omega$ but by taking the wrong temperature. Therefore, we find for the full emitted power in the configuration as shown in Fig.~\ref{Fig:Sketch} the expression
\begin{equation}
\begin{split}
  P_\omega &= P_\omega^\rp + P_\omega^{\rm sc} + P^{\rm extra}_\omega \\
           &= k_0^3 \bigl(  \Theta(T_\rp) - \Theta(T_\rs) \bigr)  \Im(\tilde{\alpha}) \biggl[ \frac{I_{xx}}{|\tilde{D}_{xx}|^2} +   \frac{I_{zz}}{|\tilde{D}_{zz}|^2} \biggr].
\end{split}
\label{Eq:PFinal}
\end{equation}
Hence, the scattered contribution $ P_\omega^{\rm sc}$ does not contribute at all if the environment of the nanoparticle is in equilibrium. Furthermore, the above expressions clearly shows that there is no emitted power if $T_\rs = T_\rp$ as expected in global equilibrium. {Note, that neglecting the radiation-correction term would lead to a nonzero heat flux in global thermal equiblibrium, because the vacuum contribution in $P_\omega^{\rm sc}$ is in this case not cancelled by the radiation-correction term in $P^{\rm extra}_\omega$. Hence, it is important to retain the radiation-correction term for having a thermodynamically consistent result.} For experiments like TINS where the sample is not additionally heated one would therefore expect only contributions which are directly emitted by the tip. On the other hand, in the experiments of De Wilde in Refs.~\cite{DeWildeEtAl2006,BabutyEtAl2013} the tip is held at the same temperature as the environment, but the samples are heated. Therfore in this experiment also the scattered contribution has to be taken into account in general. For the SNoiM in Refs.~\cite{KajiharaEtAl2010,WengEtAl2018} the tip and the sample are at ambient temperature but the surrounding  radiation field of both sample and tip is cooled down so that the here made calculations cannot directly be applied.

\section{Numerical results}

\begin{figure}
  \epsfig{file = 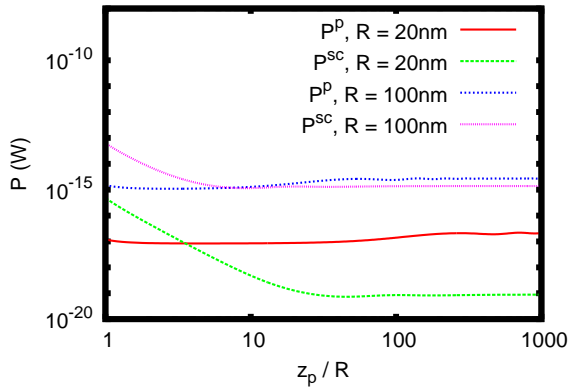, width = 0.45\textwidth}
  \caption{\label{Fig:DistanceDep} Log-log plot of the full emitted power $P^{\rm p}$ and $P^{\rm sc}$ for $T_\rp = 700\,{\rm K}$ and $T_\rs = 300\,{\rm K}$ as a function of the distance $z_\rp$ normalized to the radius $R$ for different radii $R = 20\,{\rm nm}$ and $100\,{\rm nm}$. The nanoparticle is assumed to be made of Si and the sample is made of SiC.}
\end{figure}

Let us first discuss the distance dependence of the emitted and scattered power $P^{\rm p}$ and $P^{\rm sc}$ as can be obtained by frequency integration of the spectral powers in Eq.~(\ref{eq:Pnaked}) and (\ref{eq:Pscat}). For the permittivity of the nanoparticle we take the values for Si from Ref.~\cite{Chandler2005}, and for the sample we use the permittivity of SiC from Ref.~\cite{Palik} since in the TINS experiments of Raschke a Si AFM tip has been used to study a SiC sample. As can be seen in Fig.~\ref{Fig:DistanceDep} the distance dependence for $P^{\rm p}$ and $P^{\rm sc}$ are in general quite different. $P^{\rm p}$ shows significant oscillations for distances $z_\rp$ on the order of $\lambda_{\rm th}$. When making the distance smaller the value of $P^{\rm p}$ first drops and then increases again for distances $z_\rp$ just before contact with the sample. The variation of $P^{\rm p}$ as function of distance is very small. On the other hand, for $P^{\rm sc}$ the oscillations at $z_\rp \approx \lambda_{\rm th}$ are not so well pronounced and for distances $z_\rp$ below $\lambda_{\rm th}$ the value of $P^{\rm sc}$ increases monotonically like $1/z_\rp^3$. Therefore the variation of $P^{\rm sc}$ as function of distance is large and a clear near-field enhancement is observable. Furthermore, for all shown parameters the emitted power is below 1pW! In Ref.~\cite{JonesEtAl2012} a TINS signal on the order of 20-100pW is measured. Therefore, this indicates that the thermal emission does not stem from the formost part of the AFM tip which has only a radius of approximately $20\,{\rm nm}$ but from a part of the AFM tip which has at least a radius larger than $100\,{\rm nm}$. Further investigations of the impact of the tip geometry and the contribution of higher multipole moments are needed in order to clarify this point further.

\begin{figure}
  \epsfig{file = 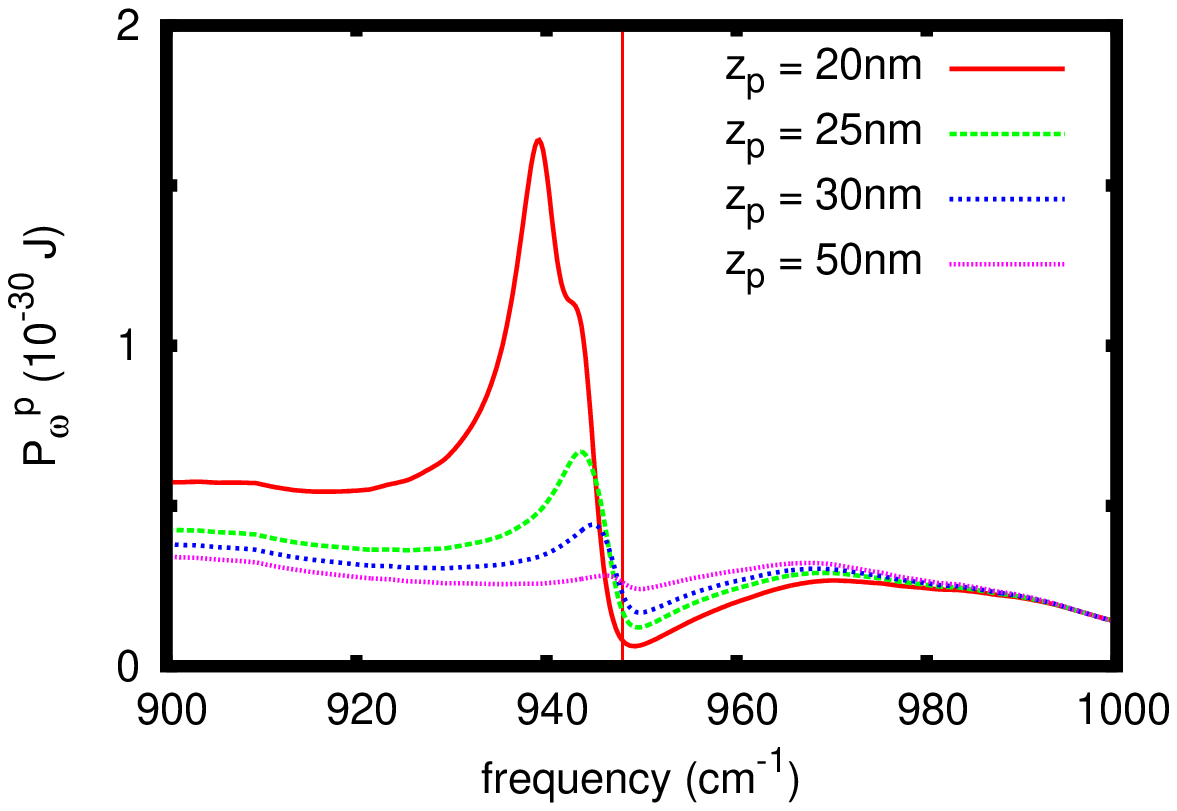, width = 0.45\textwidth}
  \epsfig{file = 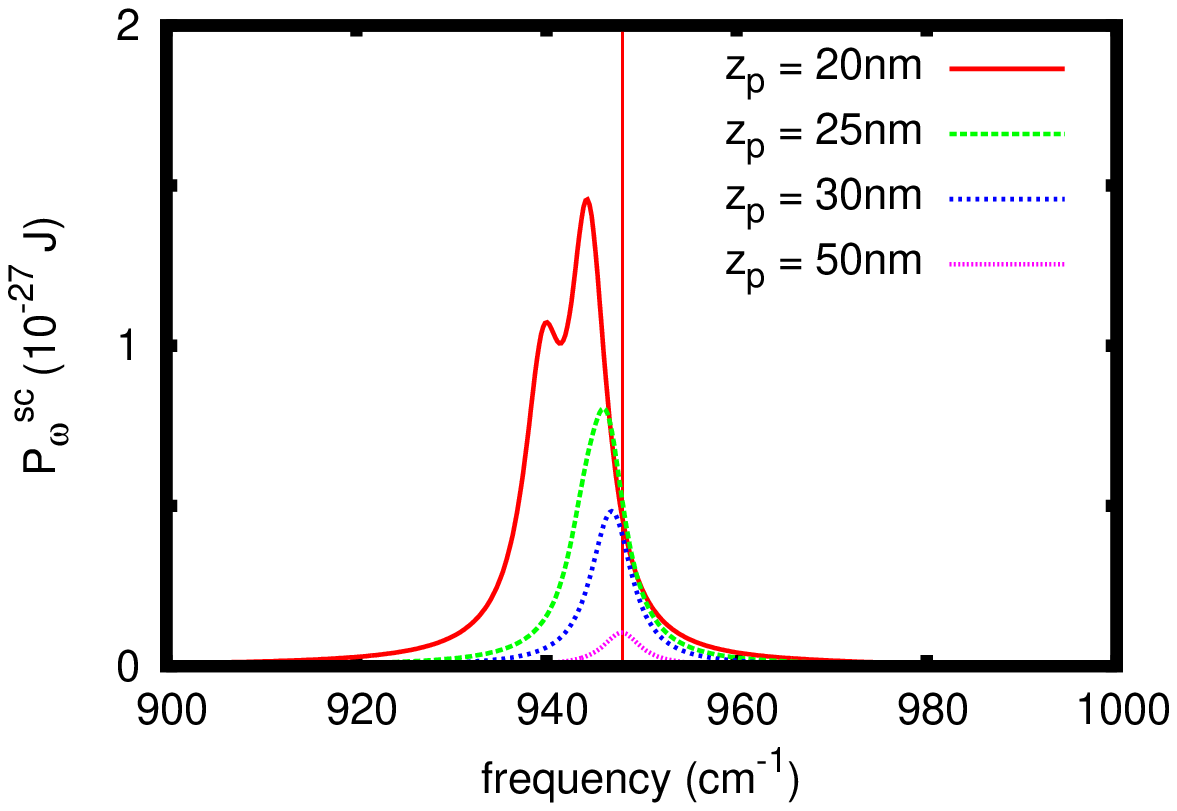, width = 0.45\textwidth}
  \caption{\label{Fig:SpecR20nm} Plot of the spectral emitted power $P^{\rm p}_\omega$ (top) and $P^{\rm sc}_\omega$ (bottom) for $T_\rp = 700\,{\rm K}$ and $T_\rs = 300\,{\rm K}$ for different distances $z_\rp$ for a fixed radius $R = 20\,{\rm nm}$. The nanoparticle is assumed to be made of Si and the sample is made of SiC. The vertical line corresponds to the surface mode resonance of SiC at $948\,{\rm cm}^{-1}$.}
\end{figure}

\begin{figure}
  \epsfig{file = 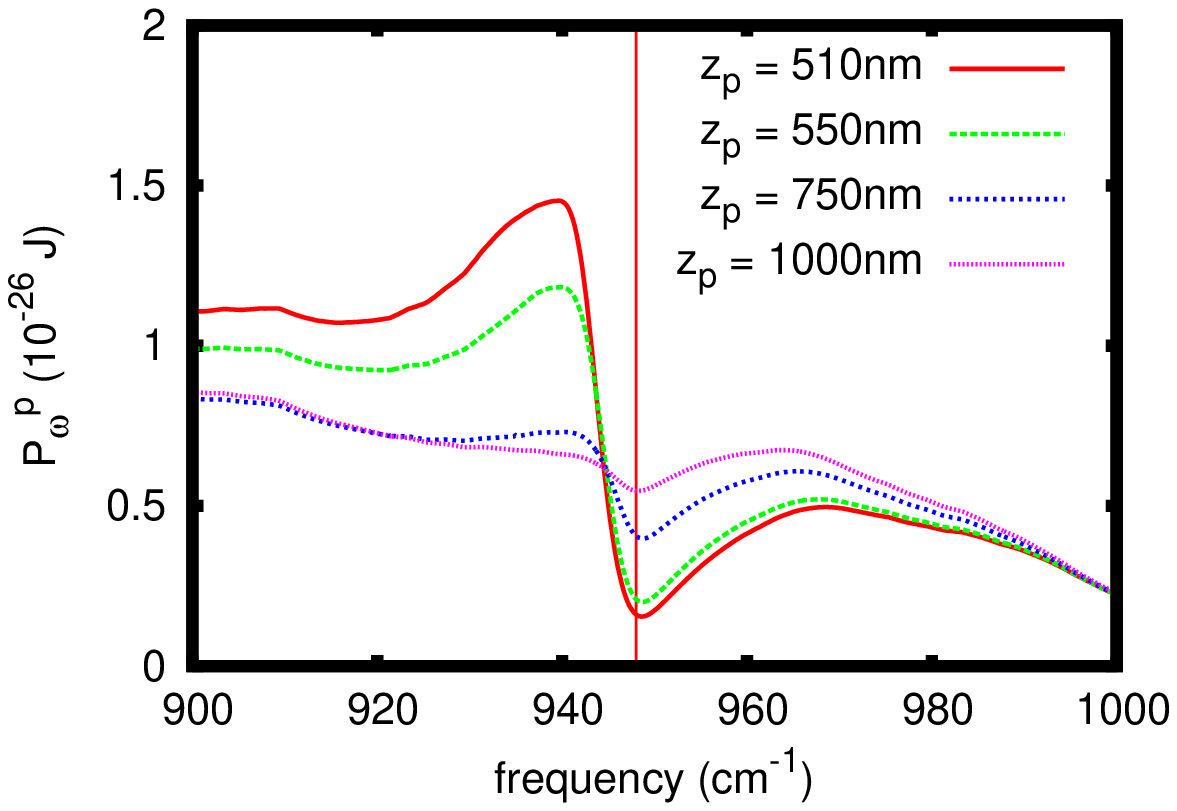, width = 0.45\textwidth}
  \epsfig{file = 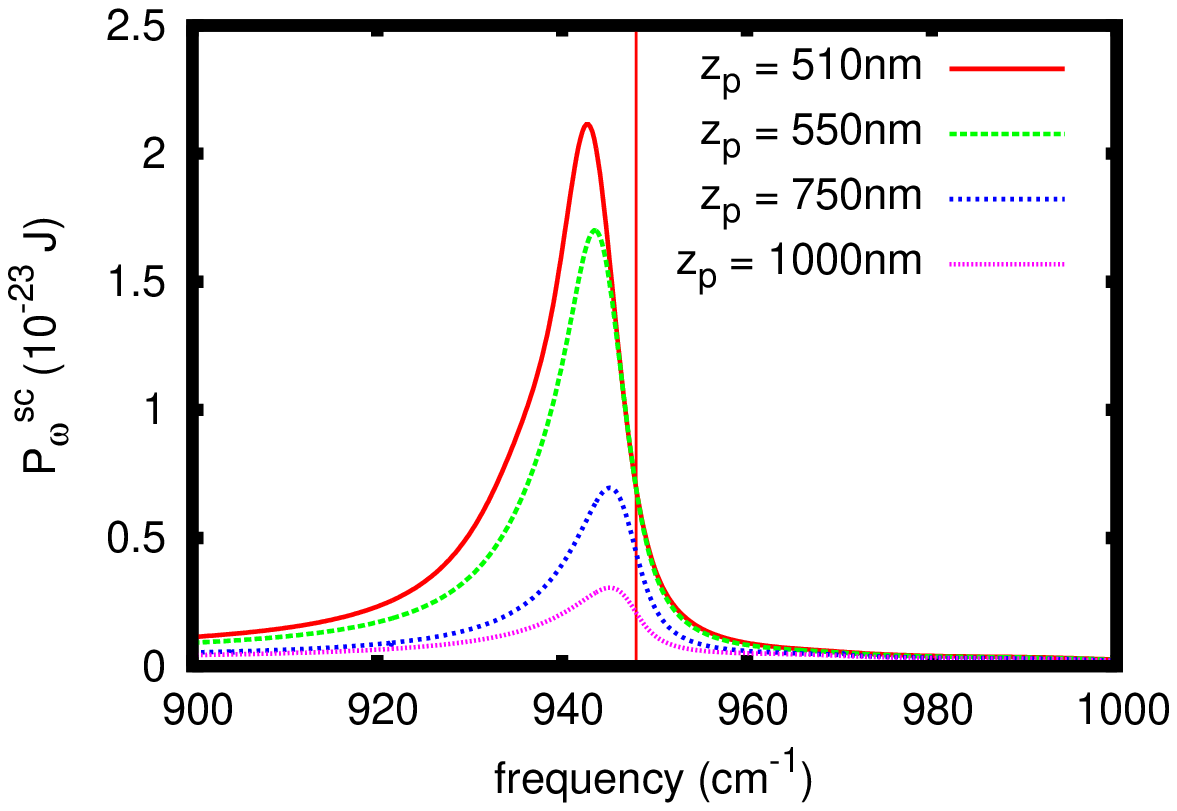, width = 0.45\textwidth}
  \caption{\label{Fig:SpecR500nm} As in Fig.~\ref{Fig:SpecR20nm} but with $R = 500\,{\rm nm}$.}
\end{figure}

In Figs.~\ref{Fig:SpecR20nm} and \ref{Fig:SpecR500nm} we show some numerical results for $P^{\rm p}_\omega$ and $P^{\rm sc}_\omega$ for $R = 20\,{\rm nm}$ and $R = 500\,{\rm nm}$. It can be seen that both quantities $P^{\rm p}_\omega$ and $P^{\rm sc}_\omega$ show a redshift of the surface mode resonance frequency of the SiC sample at  $948\,{\rm cm}^{-1}$. This redshift can be easily about $10\,{\rm cm}^{-1}$ and when considering a more spheroidal shape of the particle it can also be larger as shown for $P^{\rm sc}_\omega$ in Ref.~\cite{JarzembskiEtAl2017}. Therefore, the fact that the TINS signal of Ref.~\cite{CallahanEtAl2014} spectra which show such a redshift could be fitted is no proof that the measured signal is due to the scattering part.  The only argument in favor of  $P^{\rm sc}_\omega$ is that in Ref.~\cite{CallahanEtAl2014} a large increasement of the signal for distances below 2 microns is measured. $P^{\rm p}_\omega$ does not show such a large increasement. Actually, the signal of $P^{\rm p}_\omega$ first drops when decreasing the distance and then slighly increases for very small distances $z_\rp \approx R$.

\section{Summary and Conclusions}

We find that for a TINS measurement where only the AFM tip itself is heated the contribution of the scattered near-field radiation is exactly zero and the signal should be that of the directly emitted part. Nonetheless, the properties of the sample are included in the thermal radiation of the heated nanoparticle into the far field due to the near-field interaction with the interface. When heating up the sample as well, then we expect to have contributions of the scattered $P^{\rm sc}$ and directly radiated part $P^{\rm p}$. Our simple calculations indicate that the tip size which contributes to the TINS signal must be much larger than $R = 100\,{\rm nm}$ in order to obtain signals on the order of several pW.  Both spectra $P^{\rm p}_\omega$ and $P^{\rm sc}_\omega$ can explain the red-shift measured in TINS, whereas an enhancement of the TINS signal for relatively large distances can only be explained by $P^{\rm sc}_\omega$. Hence, these results indicate that the interaction volume in TINS is much larger than the foremost part of the AFM of approximately $50\,{\rm nm}$ \cite{CallahanEtAl2014}. Further investigations which take into account the actual geometry of the used AFM tips are necessary to have a clear understanding of the signal measured by TINS. 

To summarize, we have determined the overall thermal heat flux of a nanoparticle above a planar surface which is heated or cooled with respect to its environment within the dipole approximation. We have determined the analytical expression for the directly emitted mean power $\langle P^{\rm p} \rangle$ of the nanoparticle and the scattered power $\langle P^{\rm sc} \rangle$ together with the contribution $\langle P^{\rm extra} \rangle$. We find that a previously found expression $\langle P^{\rm p} \rangle$ is erraneous. In the case where the surrounding of the particle is in equilibrium at one temperature then $\langle P^{\rm p} \rangle$ contributes only. 

Hence, for experimental setups like TINS where a hot AFM tip is used to scan sample surfaces one would expect that the signal is due to the direct emission of the AFM tip and not due to the scattered thermal near-field radiation. In those TINS and TRSTM experiments where the sample is heated $\langle P^{\rm sc} \rangle$ will also contribute and might give a dominant contribution if the size of the tips which scatter the thermal near-field into far-field is comparable to the thermal wavelength. Our spectral results for the emitted and scattered power show that both the directly and the scattered power show a spectral shift of the near-field resonance of the phonon-polariton modes of a SiC sample as has been found in the TINS and TRSTM experiments so that from the spectra alone it is not evident if the experiments measure $P^{\rm sc}_\omega$ or $P^{\rm p}_\omega$ or a combination of both. Further investigations on the impact of the tip size and shape are necessary to shed more light on the signal measured in TINS and TRSTM.

\acknowledgments

S.-A.\ B. acknowledges support from Heisenberg Programme of the Deutsche Forschungsgemeinschaft (DFG, German Research Foundation) under the project No. 404073166.

\end{document}